\documentclass[conference]{IEEEtran}
\IEEEoverridecommandlockouts
\usepackage{amsmath,amssymb,amsfonts}
\usepackage{algorithmic}
\usepackage{graphicx}
\usepackage[noadjust]{cite}
\def\BibTeX{{\rm B\kern-.05em{\sc i\kern-.025em b}\kern-.08em
    T\kern-.1667em\lower.7ex\hbox{E}\kern-.125emX}}
\begin{document}

\title{Implementing Active Learning in Cybersecurity: \\Detecting Anomalies in Redacted Emails

\thanks{Funding source anonymized for blind review}
}

\author{
\IEEEauthorblockN{
Mu-Huan (Miles) Chung\textsuperscript{a}, 
Lu Wang\textsuperscript{a}, 
Sharon (Siyuan) Li\textsuperscript{a}\\ 
Yuhong (Alisha) Yang\textsuperscript{c}, 
Calvin Giang\textsuperscript{c}, Khilan Jerath\textsuperscript{c}, Abhay Raman\textsuperscript{c} \\ David Lie\textsuperscript{b}, Mark Chignell\textsuperscript{a}}
\IEEEauthorblockA{
\textsuperscript{a}Mechanical and Industrial Engineering, University of Toronto\\
\textsuperscript{b}Electrical and Computer Engineering, University of Toronto\\
\textsuperscript{c} Enterprise Services, Sun Life Financial Inc. 
}
}

\maketitle

\begin{abstract}
Research on email anomaly detection has typically relied on specially prepared datasets that may not adequately reflect the type of data that occurs in industry settings. In our research, at a major financial services company, privacy concerns prevented inspection of the bodies of emails and attachment details (although subject headings and attachment filenames were available). This made labeling possible anomalies in the resulting redacted emails more difficult. Another source of difficulty is the high volume of emails combined with the scarcity of resources making machine learning (ML) a necessity, but also creating a need for more efficient human training of ML models. Active learning (AL) has been proposed as a way to make human training of ML models more efficient. However, the implementation of Active Learning methods is a human-centered AI challenge due to potential human analyst uncertainty, and the labeling task can be further complicated in domains such as the cybersecurity domain (or healthcare, aviation, etc.) where mistakes in labeling can have highly adverse consequences. In this paper we present research results concerning the application of Active Learning to anomaly detection in redacted emails, comparing the utility of different methods for implementing active learning in this context. We evaluate different AL strategies and their impact on resulting model performance. We also examine how ratings of confidence that experts have in their labels can inform AL. The results obtained are discussed in terms of their implications for AL methodology and for the role of experts in model-assisted email anomaly screening.
\end{abstract}

\begin{IEEEkeywords}
Machine Learning, Active Learning, Human-Computer Interaction, Email Exfiltration, Insider Threat.
\end{IEEEkeywords}

\section{INTRODUCTION}
Large organizations are faced with ever-increasing cybersecurity threats from a variety of sources. Adversaries, armed with increasingly sophisticated tools, are looking for weaknesses that they can exploit. Defending against all possible threats is difficult because of the ingenuity of malicious agents and the many ways in which exploits can be carried out. Hardening the organizational perimeter is insufficient because many threats come either from insiders or malicious outsiders (or so-called masqueraders) who have obtained valid credentials through various methods such as social engineering. Benign internal users could also leak sensitive information if they are careless, and this can be a major  threat, especially when employees are working from home. Thus, organizations need to detect and prevent both intentional, and unintentional, data exfiltration processes before they can cause significant damage. 

The current state of the art in cybersecurity uses Machine learning (ML) methods to detect email anomalies. An anomaly detection model can support the detection of unwanted behavior such as outbound emails containing sensitive information. However, for ML models there is always the problem of how to obtain/generate the training set of labeled instances. In the domain of cybersecurity, where security and privacy concerns are the top priority, high-quality labeled data is difficult to acquire. There are typically a large number of sensors/sniffers creating noisy datasets that are difficult to pre-process and label. Without proper labeling, it is impossible to train supervised models and detect anomalous activities accurately. 

In addition, training and deploying a ML model in the domain of cybersecurity is hampered by the notorious "imbalanced data" problem. Since malicious activities tend to be rare or covert, it is difficult to construct a set of labeled instances that can be used to train a ML model. Training models with unbalanced sets of instances can generate a high number of false alarms. As a result, analysts may need to work through many alerts each day to determine which of the flagged instances are true anomalies.  

Active learning (AL) is an interactive learning method intended to improve the data labeling and model training process. The AL process usually requires only a small set of labeled data to train an initial learner model, which can then start learning from human feedback. The learner model queries human analysts with the cases that generate the greatest model uncertainty, with the goal of improving prediction/classification performances. In practice, AL may work better with cases that are on, or slightly removed from decision boundaries, depending on factors such as the distributional properties of the instance data. 

AL has been widely used for data labeling and has been found to be effective in reducing labeling costs \cite{b1, b2}. It can support ML data preparation and help ML models overcome difficult challenges. However, while AL has been tested and applied in a variety of non-expert labeling tasks, its performance has not been well studied with labeling tasks that require expertise. 

Expert labeling tasks can be viewed as a general labeling task, but labelers may not always be certain about the labels that they assign. For instance, in some image recognition tasks, if the resolution of an object in an image to be labeled is relatively low, it can be difficult for the labeler to assign a label. Previous studies suggested the appropriate way under such circumstances would be to collect “soft” labels \cite{b3}, such as aggregating regular labels from one or more labelers and transforming them into a probabilistic distribution, so as to determine the actual label based on the highest likelihood class \cite{b4, b5}.

Experts may not have complete confidence in the labels they are assigning, either because their knowledge in the domain is limited, or because the privacy of data does not allow them to have access to all the relevant information (as may often be the case in email anomaly detection). Can AL still work when there is relatively high labeling uncertainty, not just on the ML model side, but also on the side of the human assigning the labels? We propose the following research questions to study how to apply AL in a way that takes into account the labeling uncertainty of both experts and ML Models. We evaluate these questions in the remainder of the paper using an email anomaly detection case study):

\begin{itemize}
    \item \textbf{RQ1}: Are experts well calibrated in terms of assigning self-confidence ratings to their labeling decisions? (i.e., do model predictions improve when experts are more confident in their labeling decisions?)
    \item \textbf{RQ2}: Does training ML models with groups, rather than individuals, lead to better model performance? 
    \item \textbf{RQ3}: How well do experts agree with each other and are there individual differences in terms of how well different experts train ML models (where the quality of training is defined in terms of how well the subsequent models perform)? 
\end{itemize}

Studying these research questions should help us understand the effectiveness of AL (under the assumption that complete knowledge of the oracle is inaccessible) and the efficiency of analysts working with ML. In the research reported below, we carried out a case study concerning outbound email anomaly detection. In the remainder of the paper, we first review the field of AL. We then summarize the method, process, and result of a case study that examined the application of AL, in a challenging email anomaly detection process, where expert ratings of confidence in their assigned labels were also collected. 

\section{BACKGROUND}
Labeling cost has always been a major issue for supervised ML model training. In many domains, crowdsourcing \cite{b6,b7} helps resolve the issue by recruiting participants to label digits, images, or video captions. Captcha \cite{b8} is a well-known labeling use case that was initially implemented as a Turing test-like security check to identify if a user requesting a resource was a human being or a machine. The subsequent ReCaptcha application used Captcha login challenges to support large-scale book digitization \cite{b9}. However, it is unclear how to replicate the success of crowdsourced labeling in a domain like  cybersecurity where datasets are usually private and sensitive. 

Crowdsourcing proprietary labels leak data to unknown outsiders. In addition, the expertise-intensive property in the cybersecurity domain makes it impossible for non-expert participants to provide high-quality labels, and recruiting a large number of analysts is unrealistic because of the cost. While external contractors might also be used they will also be costly and will need to be entrusted with the organization’s sensitive data within an organization. Thus, Active Learning (AL) is attractive because it does not need many participants, and is feasible for an organization to carry out with existing personnel. 

Active Learning uses a feedback loop to overcome ML training “bottlenecks” \cite{b10}. For many tasks that are naturally easier for human brains, such as object segmentation and video annotation, AL may improve performance by incorporating human knowledge. We now consider different types of AL scenarios that can be used.  

\subsection{AL Scenarios}
AL can be employed with different scenarios based on the type of data flow and the target task, where there are two major scenarios: stream-based and pool-based. The stream-based scenario was first proposed as a sampling strategy to replace random sampling for certain tasks, under the assumption that the unlabeled data acquisition cost is low (or free) \cite{b11, b12}. The active learner can decide whether or not to request a label for an upcoming instance or to skip it, based on the distributional properties of the instance data. Since the decision is made for each instance in a stream (or in a sequence/row), the scenario is referred to as stream-based.

The pool-based scenario focuses on situations where a large set of unlabeled data can be acquired at once \cite{b10}. Pool-based AL typically starts with an initial learner model trained with a small set of labeled data (the initializer). The active learner then asks for labels for each instance in a large set of unlabeled data (the pool)  \cite{b11, b12}. Once  an instance is labeled it is removed from the pool and the AL process moves on and queries the next instance. 

Requesting one label at a time can sometimes be problematic in both stream-based and pool-based scenarios. For example, for an email anomaly detection task, requesting a label for a single email can be time-consuming in two ways. First, the single instance has to be investigated by an analyst. The investigation process might require considerable support, including, but not limited to, cross-departmental cooperation. Second, if there is a complex ML model underlying the active learner, retraining that model after each new labeled instance may be computationally expensive. Thus, an alternative  batch-mode scenario can be used to query multiple instances at once  \cite{b13,b14}.

The concept of batch-mode processing allows the active learner to sample a “batch” of uncertain instances (if using uncertainty sampling) at the same time. This may significantly reduce the time cost of the two traditional scenarios. However, the batch-based scenario sometimes leads to excessive redundancy in a batch, where there may be little to no information gain during an iteration  \cite{b15}. A potential solution is to choose diverse instances that involve different combinations of features so as to promote labeling efficiency. These instances will typically be some distance from the decision boundary \cite{b16}. Thus, “querying for diversity” is a way to force the inclusion of diverse instances, so as to reduce redundancy and improve query efficiency \cite{b17}.

\subsection{AL Querying/Sampling Strategies}
AL requires a sampling strategy, and one of the most common strategies is to prioritize instances for labeling based on the ML model uncertainty \cite{b18,b19}. Using this type of uncertainty minimization strategy the queue of instances for labeling is ordered based on their uncertainty rankings so that the model can maximize information gain after each labeling judgment. For instance, as shown in Figure 1 (binary classification), the orange bubbles closer to the decision boundary are more likely to be queried in AL because they have higher model uncertainty (due to their proximity to the decision boundary). 

\begin{figure}[h!]
    \centering
    \includegraphics[scale=0.4]{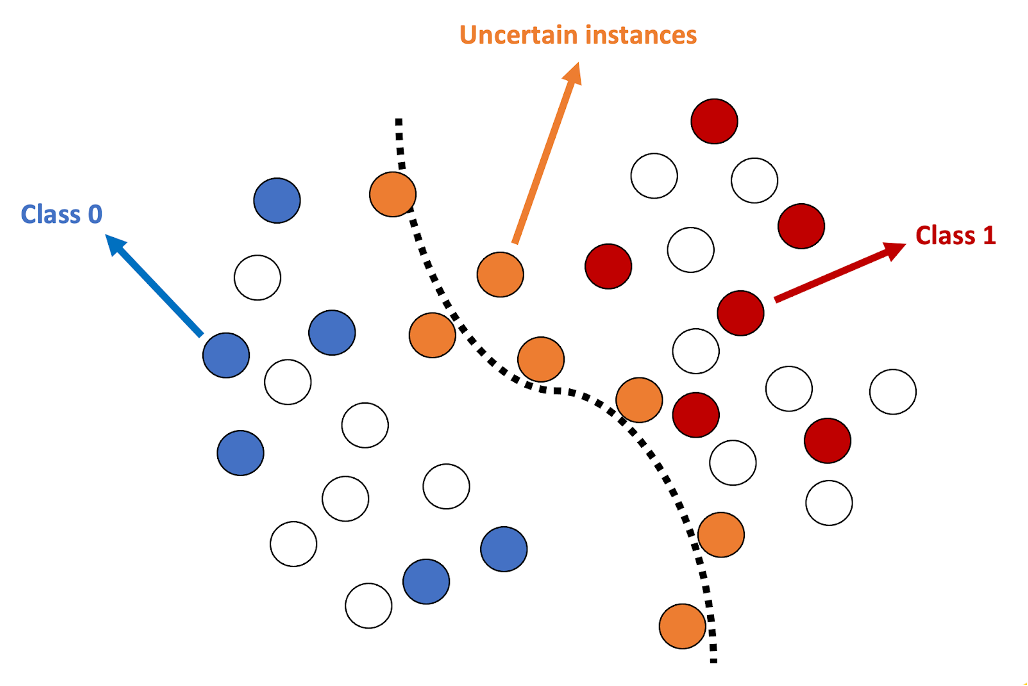}
    \caption{Active Learning implementation in a binary classification task}
    \label{}
\end{figure}

There are several ways to calculate uncertainty \cite{b20} from predicted probability. In a 3-class classification task, for example, the uncertainty can be calculated with Least Confident (eq. 1), where \textit{y'} is the most likely prediction among the 3 classes. If the probabilities of each class are [0.15, 0.65, 0.20], the uncertainty can be calculated as 0.35. 

\begin{equation*}
    L(x) = 1-P(y'|x) \tag{eq. 1}
\end{equation*}

The least confident calculation omits the information of the other 2 classes, which is not optimal for many cases. Thus, a second way of using the classification Margin (eq. 2) was proposed. This strategy focuses on the difference between the top and second likely classes having the highest probability \textit{y'} and second highest probability \textit{y"}. Under this strategy, the uncertainty of a row having predicted probabilities of each class as [0.15, 0.65, 0.20] would be calculated as 0.45.

\begin{equation*}
    M(x) = P(y'|x) - P(y"|x) \tag{eq. 2}
\end{equation*}

Another commonly used strategy to calculate uncertainty is by calculating the classification Entropy (eq. 3). Classification Entropy corresponds to the expected log-loss \cite{b20}, which is proportional to the average number of guesses one has to make to find the true class \cite{b21}.

\begin{equation*}
    H(x) = -\sum_{i} P(x_i|x) log P(x_i|x) \tag{eq. 3}
\end{equation*}

There are also other querying strategies in addition to uncertainty sampling \cite{b10}. Multiple models may be organized into a pipeline, and combinations of models may be utilized (e.g., Query-by-Committee, Disagreement Sampling, Expected Model Change, or Estimated Error Reduction, etc.) to improve model performance in each feedback loop. Thus, the decision of the querying strategy is task-oriented and subject to change based on the algorithm underlying the active learner. 

While other querying strategies (e.g., \cite{b22, b23, b24, b25}) have been proposed, the uncertainty-driven sampling strategy remains a simple and straightforward approach. In the remainder of this paper, we explore (within an applied setting) how uncertainty-driven sampling for AL can be improved by also considering the uncertainty of the human labeler. 

\section{CASE STUDY}
This case study involved a large financial services company. The effectiveness and efficiency of AL, as well as the human factors of expert-model interactions, were examined with respect to an outbound email anomaly detection task. We focused on email behaviors that violated company policy (and, in particular, emails that employees sent to their own email addresses). This type of behavior was judged to be problematic because the company required that employees worked on “locked-down” company computers. Employees might be circumventing this policy by sending email messages to themselves to covertly transfer data to other (insecure) computers that they used. 

Prior to the initiation of our research, the detection task was conducted at the company weekly, and manually, by one security analyst, who reviewed emails on a set of dashboards in order to label each email/case as anomalous or not. The dashboards had filtering functionalities, so that the analyst could screen emails to find those with particular combinations of sensitive terms. After filtering, the resulting set of email activities was visualized with two scatterplots of all email sizes and counts of all users; two trend plots of a selected user’s past email count and size histories; and a detailed table of the selected user’s email activities on that day. The whole process usually took roughly two business days (every week) to be completed, which included filtering from roughly 10,000 emails, screening and reviewing over 1000 filtered emails from about 25 to 50 different users who sent numerous and/or sizable emails that contained certain sensitive terms in subject lines or attachment names. 

There was also a second stage of investigations conducted by another team that had higher security clearance. The follow-up investigations included reviewing detailed email attributes, email bodies, attachment details, and sometimes talking to the sender’s manager. This process was not automated in any way and consequently the task of reviewing each email was time-consuming. 

The whole manual detection process was inefficient and likely prone to misses/false positives. Thus, ML interventions were required to improve detection speed and accuracy. There was a previous attempt by an outside vendor to use automatic ML (with previous manual detection outcomes as the training/testing datasets) for anomaly detection on the company emails. However, this resulted in a 96\% false alarm rate, with only  4\% of the anomalies flagged in that work proving to be true anomalies after further investigation. The outside vendor had relied on manually assigned labels provided by the company and we suspected that low labeling quality may have been one reason for the poor automated ML prediction performance in this case. This led us to consider the use of Active Learning, to exploit human knowledge more efficiently in labeling data. 

\subsection{Scope and Data}
The dataset used in this study was comprised of 27 columns describing each email, including sender, recipient, subject, attachment, attachment size, sender identifiable data (role, location, hired date, status, etc.), DateTime variables, and whether or not some sensitive terms were mentioned in the subject line or attachment name. In addition, we counted and summed the number of recipients. We also included two binary variables indicating whether or not an email contained certain sensitive terms (previously used in manual detection tasks) in its subject line (first variable) and attachment names (second variable). We also calculated the similarity of the email address and user name as a proxy for the likelihood that the person was sending an email to one of their other email counts (using Levenshtein distance \cite{b35}). This similarity was then used as an additional variable. 

There were approximately 320,000 rows in the raw, unlabeled dataset, which comprised two weeks of outbound email. This raw dataset was split into eight sets of unlabeled data, with each set being assigned to one of the eight rounds (by the order of date/time to mimic a real detection task) of Active Learning phases in this case study. There was also another pre-labeled dataset consisting of 200 instances randomly selected from the raw data. This latter data set of pre-labeled instances was used to train the initial ML detection model. 

One factor constraining the labeling task was that the data provided to the analysts did not contain email body and attachment details. That extra data was only released to a different team (not studied here) that had higher security clearance and was charged with investigating potential anomalies once they were detected. Thus, in this screening/filtering stage, human judges were forced to make labeling decisions based on subject lines, file names, and some user-identifiable information. While these variables should be enough to detect careless exfiltration, they might potentially lead to more anomalies being flagged (“just in case”) in turn leading to a higher rate of false positives). 

\subsection{Methodology}
This study implemented an Active Learning (AL) process by having a model pre-label data after human analysts provided their own labels. The model was then updated based on the labels provided by the analysts. The AL process incorporated the following components that are typically used:

\begin{itemize}
    \item A small set of pre-labeled data
    \item An AL model initialized with the pre-labeled data
    \item A selected AL scenario and a sampling strategy that is compatible with the analyst work role
    \item A termination criterion (e.g., an accuracy threshold or asymptotic improvement in accuracy)
    \item A method to evaluate the outcome
\end{itemize}

The pre-labeled data subset, consisting of 200 cases, was used to initialize the active learner (the AL model). We used a ground truth dataset prepared by another team who had access to email details. We used a batch-based scenario where each analyst labeled multiple instances in a round. The overall active learning process used in this study is summarized in Figure 2.  

\begin{figure}[h!]
    \centering
    \includegraphics[scale=0.25]{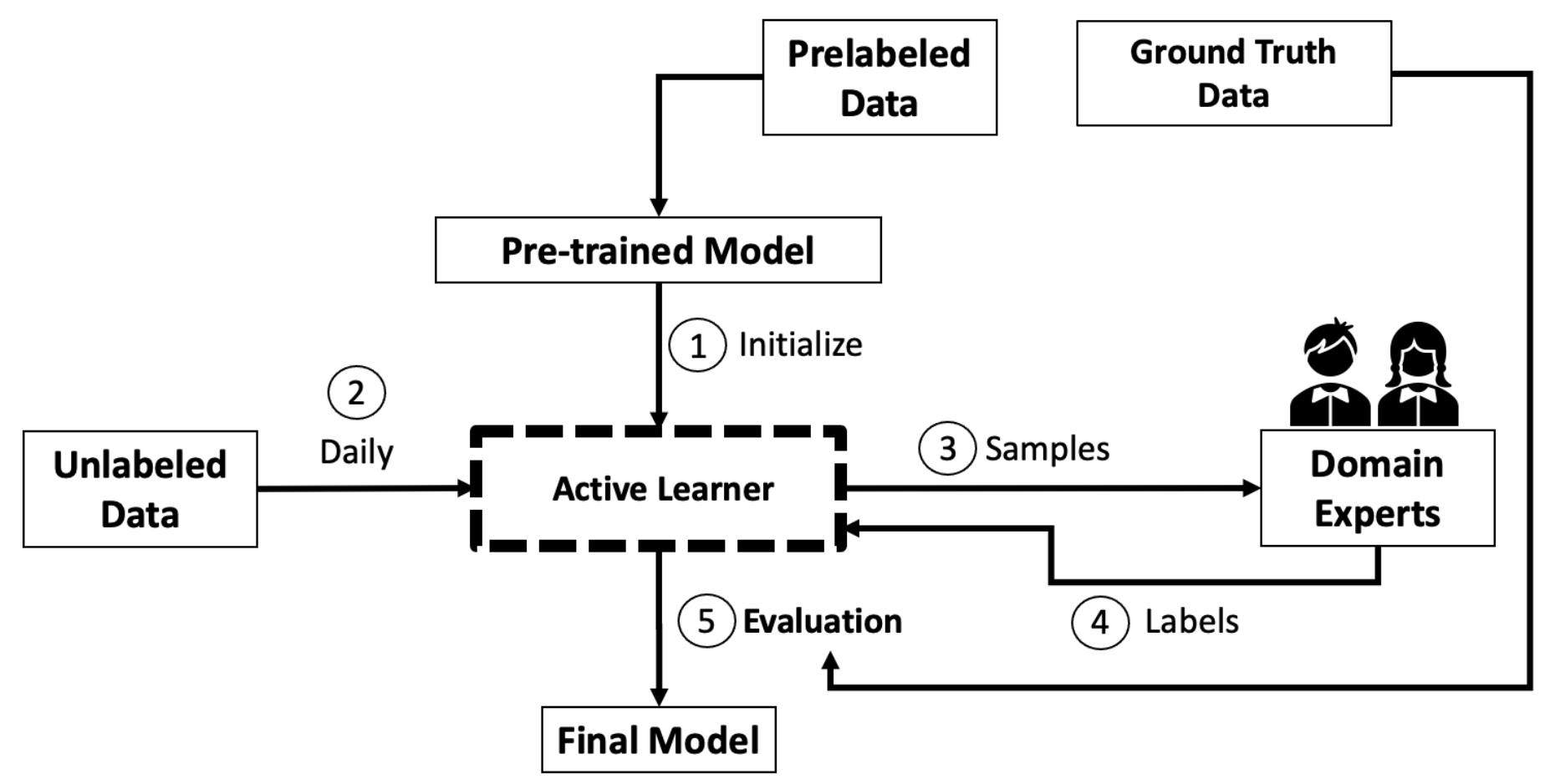}
    \caption{Overall Active Learning process used in the case study}
    \label{}
\end{figure}

One of the problems with active learning is that it can be frustrating \cite{b26} for analysts to have to label cases that the algorithm is uncertain about, since the analyst will likely be uncertain about those cases as well. Thus we can expect that active learning will work best when analysts are able to clearly label the cases that the algorithm finds ambiguous. Presumably, in practical situations, analyst labeling of ambiguous cases will be limited by how willing analysts are to carry out what could be a potentially frustrating task.  

Previous research conducted by Carcillo et al. \cite{b27} tested various AL sampling strategies with a credit card fraud detection task (which is also an AL anomaly detection task that requires expertise). They used an HRQ (highest risk query) strategy where instances with the highest sensitivity/risk were prioritized for labeling. 

We based our methodology on that described by Carcillo et al. but with adjustments to meet the needs of our case study. We used a high-risk query (HRQ) approach where the analysts were shown a high proportion of expected anomalies. However, we also included a small number of uncertain and random instances in each feedback loop for exploratory purposes. One reason for using the HRQ approach was that it may be more motivating for human analysts when they are reviewing a significant number of anomalies. 

\subsubsection{Participants}
There were 10 participants in the experiment, representing all the company employees who had the required expertise in the work group that we collaborated with. Participants performed the labeling tasks as part of their work duties. The data resulting from the experiment was collected by a third party within the same company who had no mutual interest nor conflict with the participants. A set of 10 anonymized IDs were generated and assigned to the participants by the third party. Throughout the experiment, the researchers only had access to the IDs and labels, so they were unable to link model performances with specific participants. In contrast, the third party could link IDs to participants, but could not tell which model was trained by which participant. 

After completing the task the participants were given a choice of whether or not they wished to release their anonymized data for research use. This decision was anonymous and there were no repercussions for people choosing not to complete their labels, or who did not make their data available for research use. This procedure was approved by the University of X (name omitted for blind review) ethics review board (Ethics protocol number XXXXX - number omitted for blind review). 

The labeling task was carried out as part of participants’ work practice and there were 10 participants. For each email instance to be labeled, the task was to review the data provided (subject lines, size, file names, etc.), and then determine if the evidence warranted further investigation due to possible data exfiltration activity (inappropriate transfer of confidential data). The participant then provided a label for the instance (anomalous or not anomalous) and moved on to the next email instance in the queue. We used LightGBM \cite{b28} as the underlying active learner model in this case study, for its efficiency and performance in classification tasks. Similar to the strategy successfully used by Carcillo et al. \cite{b27} we used high risk queries (i.e., instances that the ML prediction model was relatively certain were anomalies). 

During each round (day) in the study, each participant labeled 20 instances including 14 high-risk queries (HRQs), 3 uncertainty queries (UQs), and 3 random queries (RQs) each day over 8 business days. The amount of data that could be collected was constrained by the amount of time that supervisors at the company thought should be devoted to this work. The raw data was split into 8 unlabeled datasets, so as to mimic a realistic process of daily labeling/detecting tasks. There were no stopping criteria in this case study because we wanted to focus more on human interactions, rather than looking for a certain level of model performance. 

\subsubsection{Study Design}
Our case study focused on the following constructs/factors:

\begin{itemize}
    \item Individual differences (in terms of experience, expertise, etc.)
    \item Collaboration effect (when multiple analysts train one model)
    \item Human uncertainty (i.e., the analyst self-reported confidence level while providing labels, and whether this affects the active learning process)
    \item Human factors problems noted by analysts during their interaction with the AL system
\end{itemize}

To evaluate possible human factors concerns, as well as answer our proposed research questions, we separated participants into 3 teams (also shown in the following Table 1, where the different models are labeled with the letters A through F): 

\begin{itemize}
    \item \textbf{Individual}: each participant in the individual team trained (gives feedback to) their own model (B, C, and D) throughout the experiment
    \item \textbf{Swap}: each participant in the swap team trained their own model (E and F) in the first half of the experiment and switched to train a different model (F and E) in the second half of the experiment
    \item \textbf{Group}: participants in the group team trained one model (A) together - in this approach an instance was labeled as 1 (True) if two out of five analysts believed it to be 1 (True)
\end{itemize}

This experimental design allowed us to compare the performance of individuals, both amongst themselves and as compared to participants working as a group.

\begin{table}[h!]
    \centering
    \includegraphics[scale=0.4]{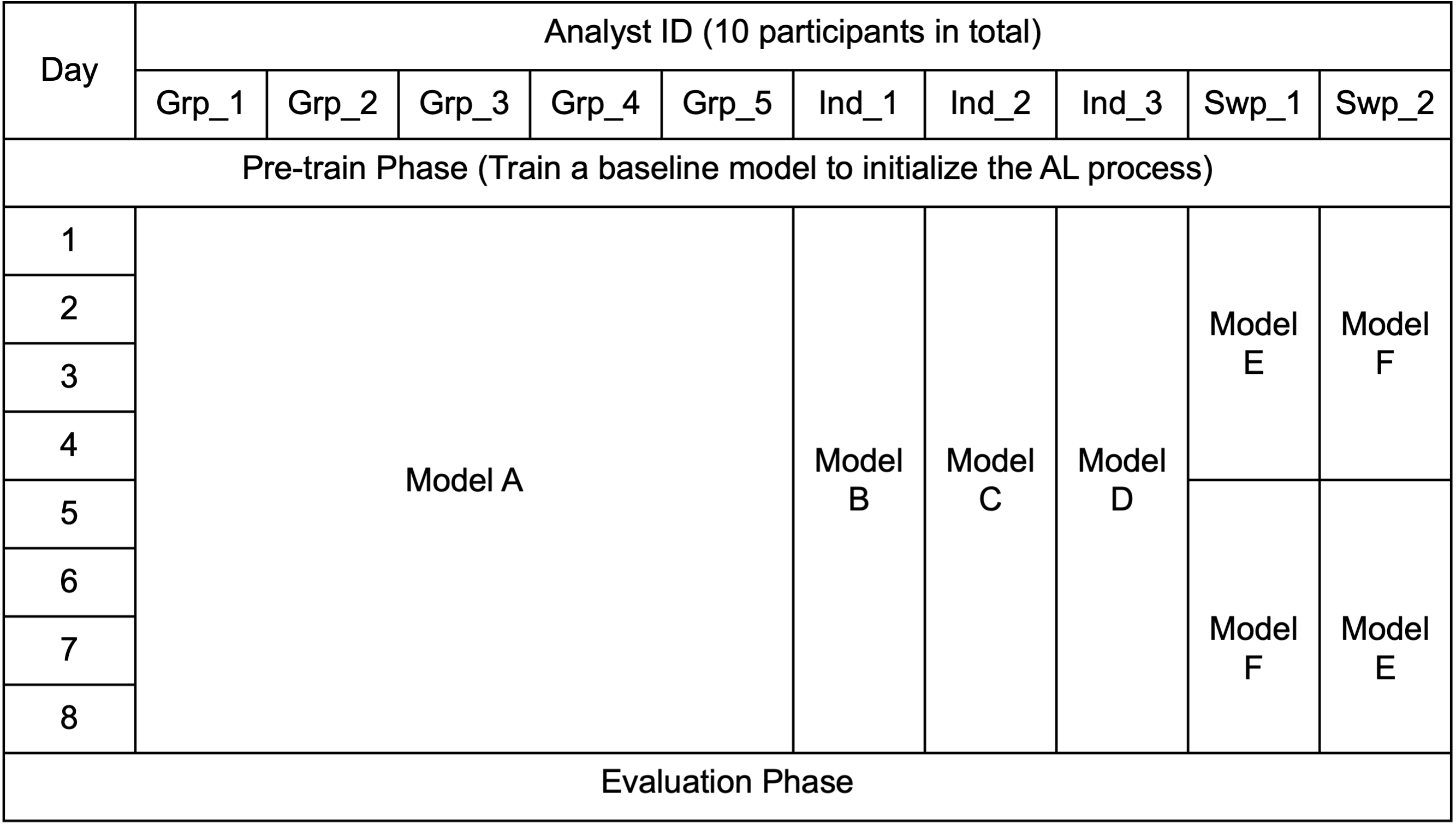}
    \caption{Planned group, swap, and individual training teams for AL experiment}
\end{table}

During the AL process, we asked participants to provide their level of confidence with respect to each label provided. This not only allowed us to examine the impact of label uncertainty but also provided a way to train additional (multiclass) models. The self-reported confidence levels were integers ranging between 0 to 10 (with 0 representing no confidence and 10 representing complete confidence). The average confidence levels and their variations for each class are shown in Figure 3. 

\begin{figure}[h!]
    \centering
    \includegraphics[scale=0.5]{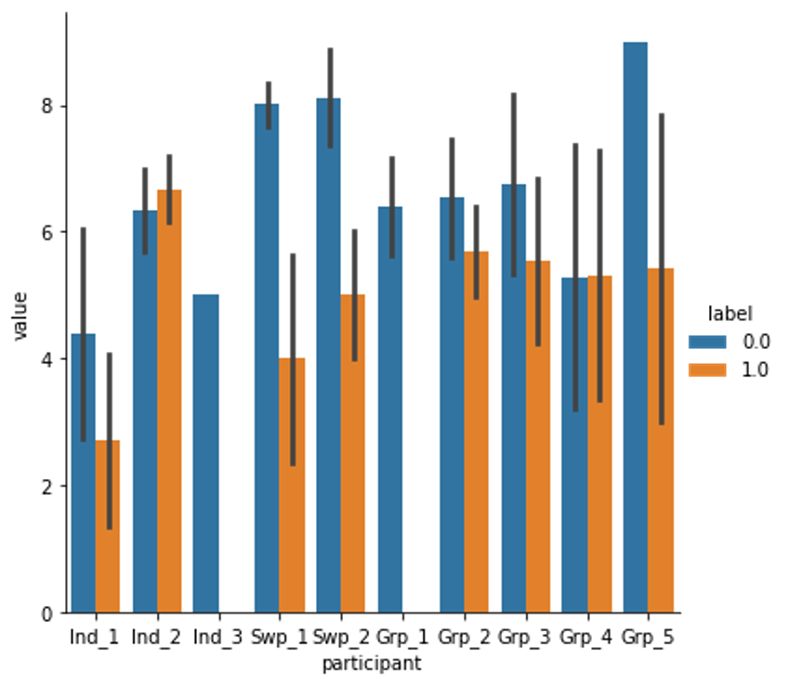}
    \caption{Pre-labeled dataset participant confidence values’ means and error bars for each label class}
    \label{}
\end{figure}

As can be seen in Figure 3, confidence levels were generally higher for 0 (False) instances compared to 1 (True) instances. Participants were much less confident when assigning 1 (True) labels. Thus, a transformation, as shown in the following equation 4, was applied to convert confidence into a pseudo probability of the instance being 1 (True). 

\begin{equation}
    \begin{cases}
      10 - Conf_i , \forall Label_i=0\\
      ROUND((Conf_i+10)/2) , \forall Label_i=1
    \end{cases} 
    \tag{4}
\end{equation}

We based our methodology of data collection and transformation on that described previously by Méndez Méndez et al. They found that collecting participant confidence levels and transforming them into probabilities of true positives (in video annotation tasks) would be a better way for data annotation, compared with only collecting binary labels \cite{b3}.

\subsection{Evaluation Schemes and Metrics}
In the last phase of the experiment, we needed a method to evaluate AL results and analyst-model interactions. We used two different ways to evaluate model performance:

\begin{enumerate}[]
    \item Typical ML model evaluation metrics such as precision and recall values\\
    
    Since the dataset is highly imbalanced, we also presented AUPRC (area under the precision-recall curve), which is preferred over AUROC (area under receiver operating characteristic curve) for its better representation of model performance with imbalanced datasets \cite{b29}. We also included the F-beta score, to highlight the tradeoff between preferences of performance favoring more on either precision or recall of a model. The test dataset was prepared by another investigation team who had full access to read the email details, including the email body and attachment file content. \\

    \item Amount of expert agreement with the model\\
    
    Agreement with the model was determined by assessing the extent to which the 14 HRQ cases were detected as anomalies in each round of the task.

\end{enumerate}

In addition, a questionnaire was prepared at the end of the experiment. This questionnaire interviewed participants regarding their opinions and experiences with the AL process, so as to help us understand better in terms of the human factors and the interactions during AL.

\section{RESULTS}
\subsection{Model Performance}
In this section, we show the results for AL with domain analysts. We begin by examining model performance and then address the research questions. Since the duration of AL in this study was relatively short due to the limited availability of domain analysts, we focus in our analysis on trends observed in the data and on lessons learned for implementing future AL systems in this type of context. 

\begin{figure*}[h!]
    \centering
    \includegraphics[width=\textwidth]{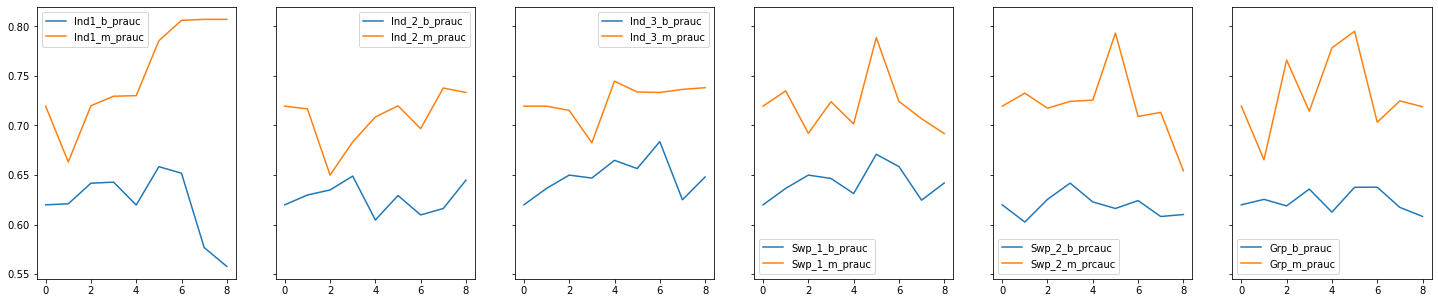}
    \caption{Binary and multiclass AL model performances of each team in AUPRC}
    \label{}
\end{figure*}

Binary model performance (shown in the blue lines) was consistently worse than multiclass model performance (orange lines) throughout the study, as shown in Figure 4. Binary labeling (as compared with multiclass labeling) has also performed poorly in other studies (e.g. crowdsourcing studies by \cite{b30, b3}).  It may be unreasonable to expect clearcut binary judgments in ML tasks where experts experience high levels of uncertainty in making the judgment.  

During training, true anomalies that were labeled by experts as non-anomalies, with confidence ratings of less than 5, were reclassified as true positives, as has been done elsewhere (e.g., \cite{b3}). The resulting multiclass models (shown in the orange lines in Figure 4) performed better than the binary models, likely due to the classes being more balanced during evaluation. When comparing multiclass models, the models trained by the swap and group participants did not outperform models trained by individual participants. Detailed performance metrics for the different participant combinations are shown in Table 2.

\begin{table}[h!]
    \centering
    \includegraphics[scale=0.45]{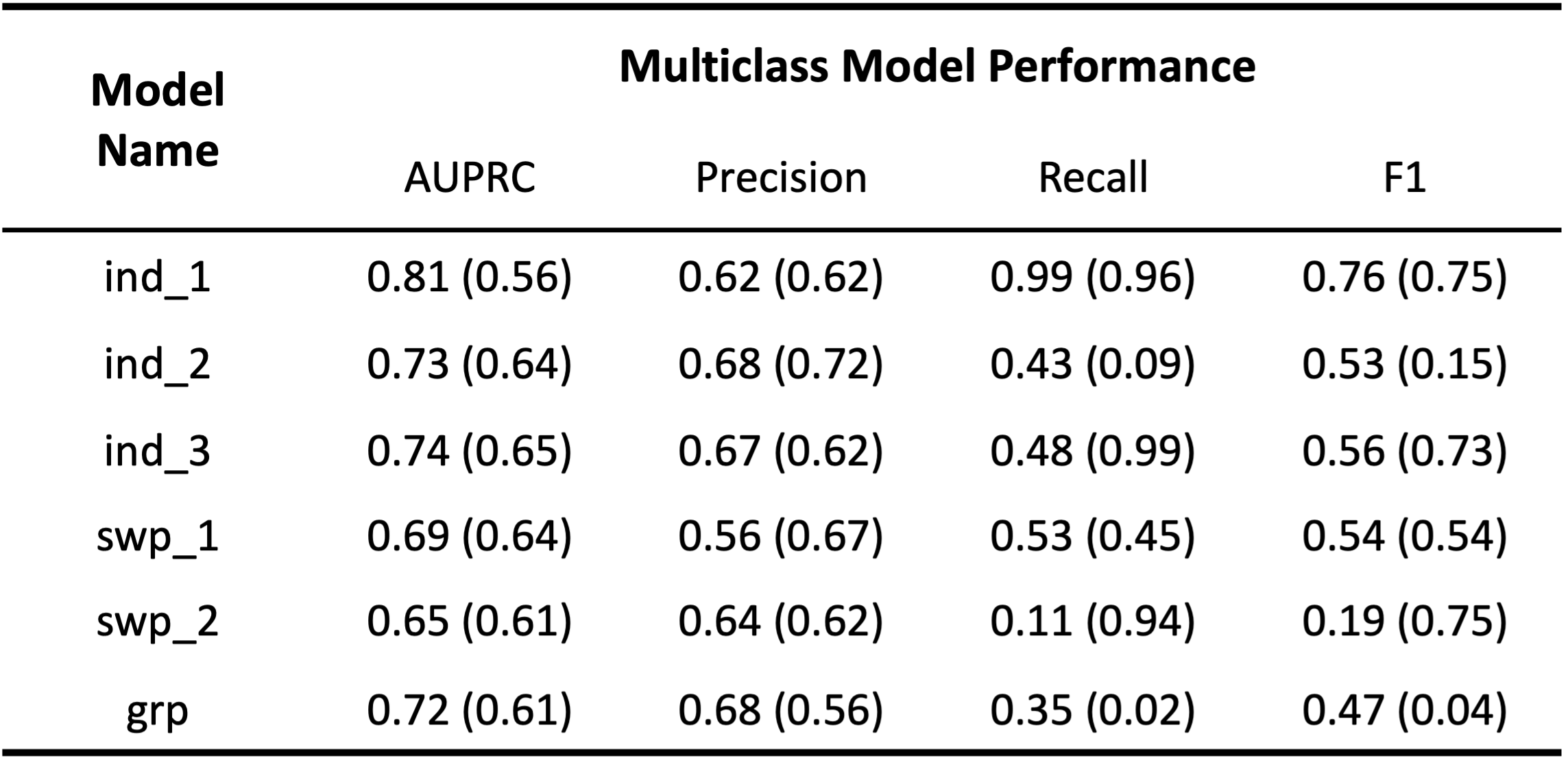}
    \caption{AL performances of multiclass and binary models (shown in parentheses)}
\end{table}

We also implemented evaluation using a second method described in section 3.3 (evaluation method B). In this latter approach, we calculated the percentage of the HRQs being labeled as 1 (True) throughout the case study. This evaluation was possible because the 14 HRQ instances of each round were the top anomalies predicted by the AL model. We then assessed whether or not these predictions matched analysts' decisions/labels in order to assess model performance.

As shown in Figure 5, while the average proportion of “True” labels assigned by participants in early rounds was low, an upward trend can  be observed (see Figure 5). The statistical significance of that trend was confirmed using a (non-parametric) Jonckheere Terpstra trend test \cite{b31, b32}, demonstrating a statistically significant trend (p=0.02), as shown in Table 3. This result showed that AL models trained with expert-supplied labels were getting more “True” labels over time. As discussed earlier the 14 HRQ instances were model-predicted top anomalies. Thus the positive relationship here represented the improvement of model accuracy in terms of matching labeling predictions to expert judgments.

\begin{figure}[h!]
    \includegraphics[scale=0.5]{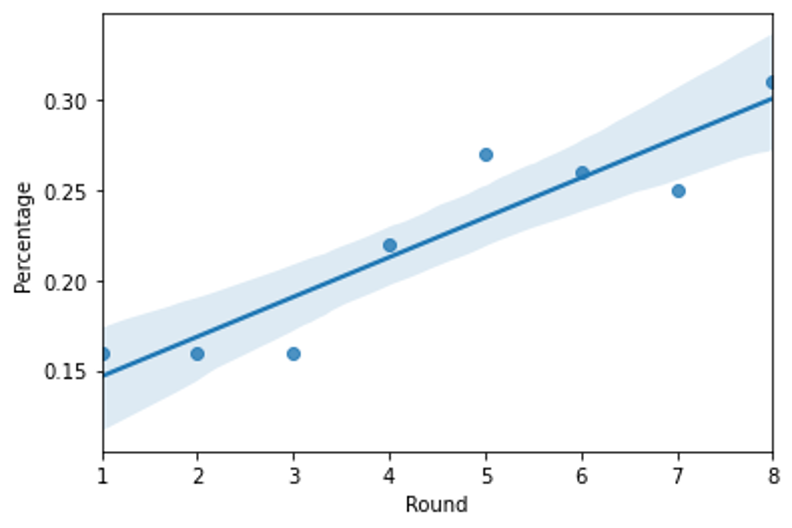}
    \caption{Trend of average 1 (True) label percentages of the 14 HRQ instances in each round}
    \label{}
\end{figure}

\begin{table}[h!]
    \centering
    \includegraphics[scale=0.3]{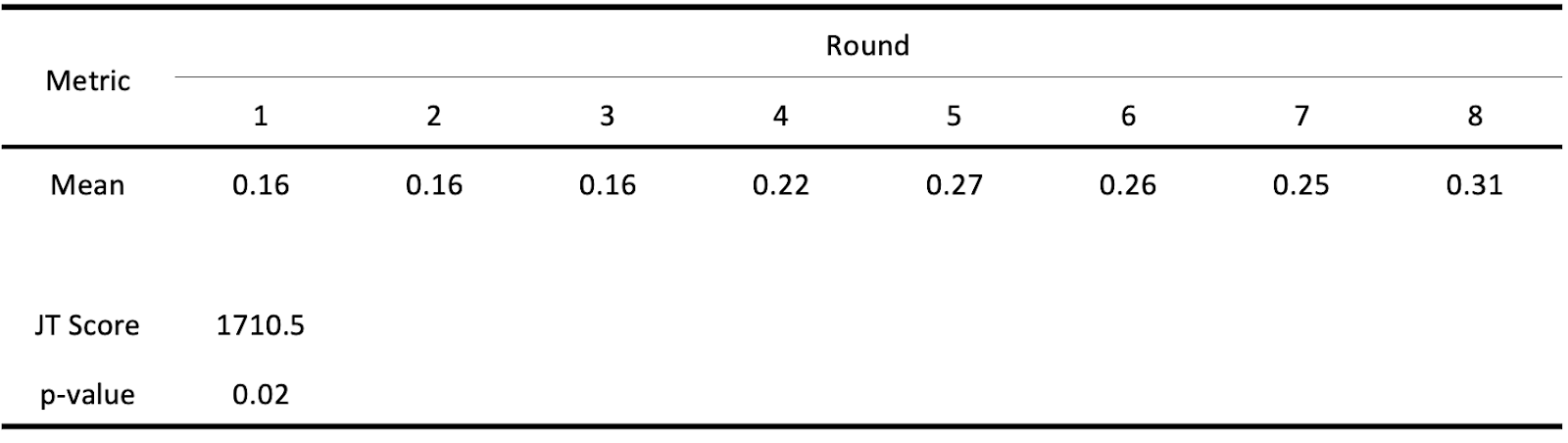}
    \caption{Jonckheere's trend test of average 1 (True) label percentages of the 14 HRQ instances in each round}
\end{table}

\subsection{Individual Difference and Label Reliability}
As shown in Figure 6 (where the error bars indicate 95\% confidence intervals), participants had differing distributions of confidence ratings with some tending to have more variable ratings or having higher or lower mean ratings, than others. These differences were observed at the outset (rating differences in the first round of the task are shown in the left panel of Figure 6), where all participants were queried using the same 20 instances. Strong individual differences were also observed over all the rounds (as shown in the right panel of Figure 6). 

\begin{figure*}[h!]
    \centering
    \includegraphics[scale=0.45]{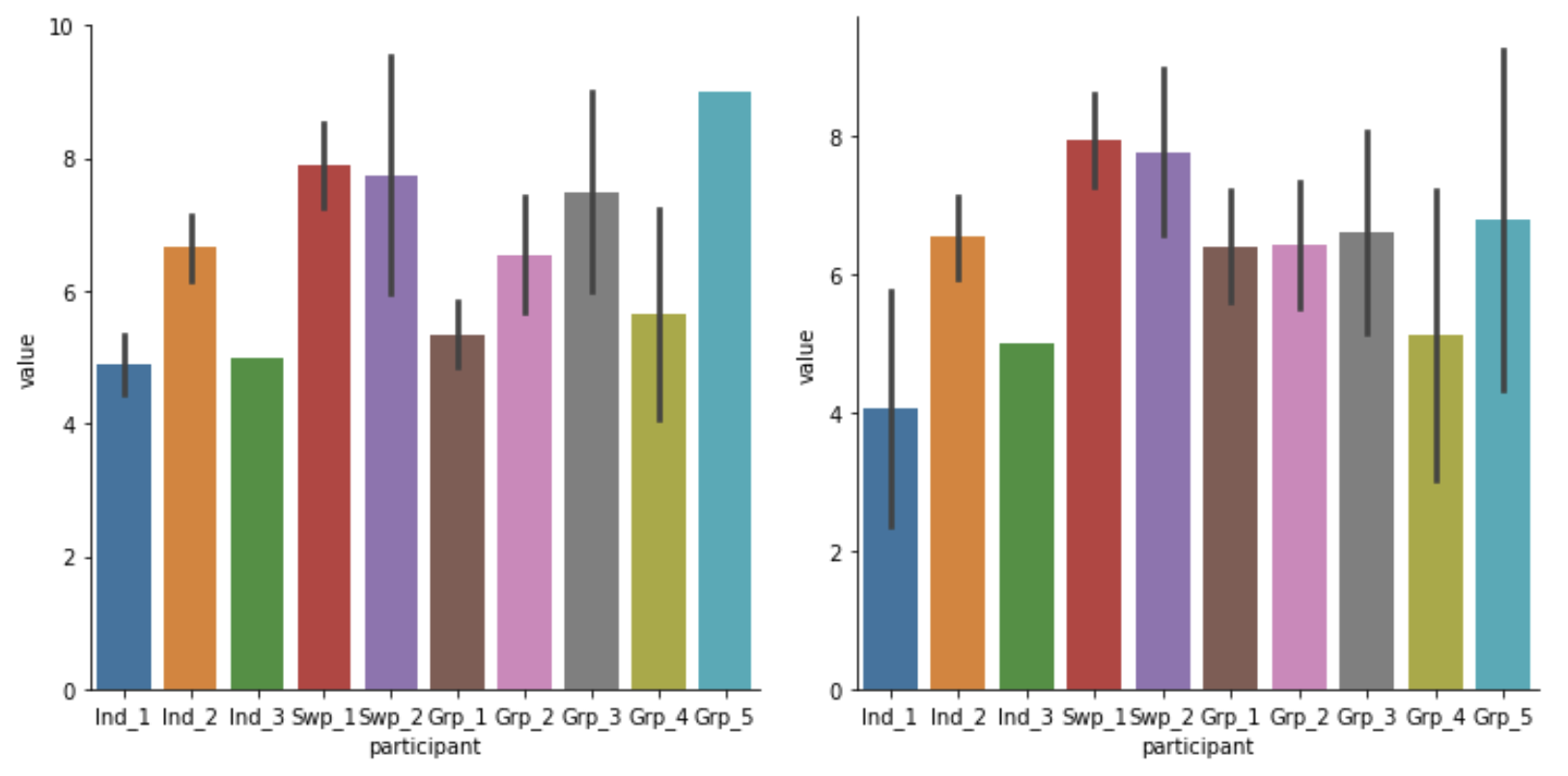}
    \caption{Individual differences in terms of average confidence level values and their variations in round 1 (left) and in all rounds (right)}
    \label{}
\end{figure*}

As a measure of internal consistency within the group, we computed Krippendorff’s alpha values \cite{b33, b34} for the decisions made each day (using the transformed multiclass labels as described earlier) within the group training team (all participants in this team labeled the same 20 instances in every round). Variations in this measure of similarity of confidence ratings can be seen in the left panel of Figure 7, with a generally downward trend in the values being observable. This indicates that analysts in the group training team disagreed with each other more (not less) over time. This may reflect the fact that there was no feedback or consultation between the participants during the study and that motivation may have varied more for some participants than for others, leading to less agreement over time. 

\begin{figure}[h!]
    \centering
    \includegraphics[scale=0.45]{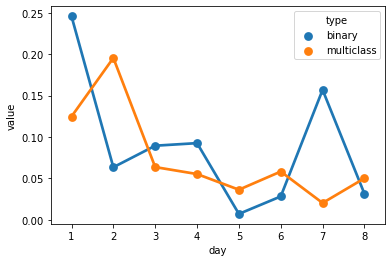}
    \caption{Krippendorff’s alpha values for both binary and multiclass models trained with group participants}
    \label{}
\end{figure}

Regardless of its cause, the observed increase in disagreement between experts in the group likely led to reduced label quality, and worsening performance of the group training model as compared to the performance of models trained by individual participants. Based on these results we suggest that there should be one or more training sessions with group participants to attain consensus (in practical situations where this technique is to be used). Participants can also be screened based on the results of preliminary labeling of cases, with subsequent AL focusing on high-performing individuals or small subgroups of high-performing individuals. This should improve the resulting AL model performance relative to current AL practice where groups of judges are used that are heterogeneous with respect to labeling ability.  

\subsection{Model and Human Uncertainty}
In terms of label reliability, there exist two types of uncertainties that may affect AL model performances under the set-up of this study: model uncertainty and human uncertainty. Model uncertainty is the measure of how uncertain the model is about some instances, where the most uncertain ones to the model would be queried when using an uncertainty querying strategy; whereas human uncertainty is a latent property that underlies each analyst’s decision, affecting the confidence level as well as the label they would be giving. As shown in Table 4, the most information gain can be obtained when the human is certain about the instances being judged, whereas the model has high uncertainty for those instances. In such cases, the human can guide the model by providing correct labels, and providing high information gain by resolving the model’s uncertainty.

\begin{table}[h!]
    \centering
    \includegraphics[scale=0.35]{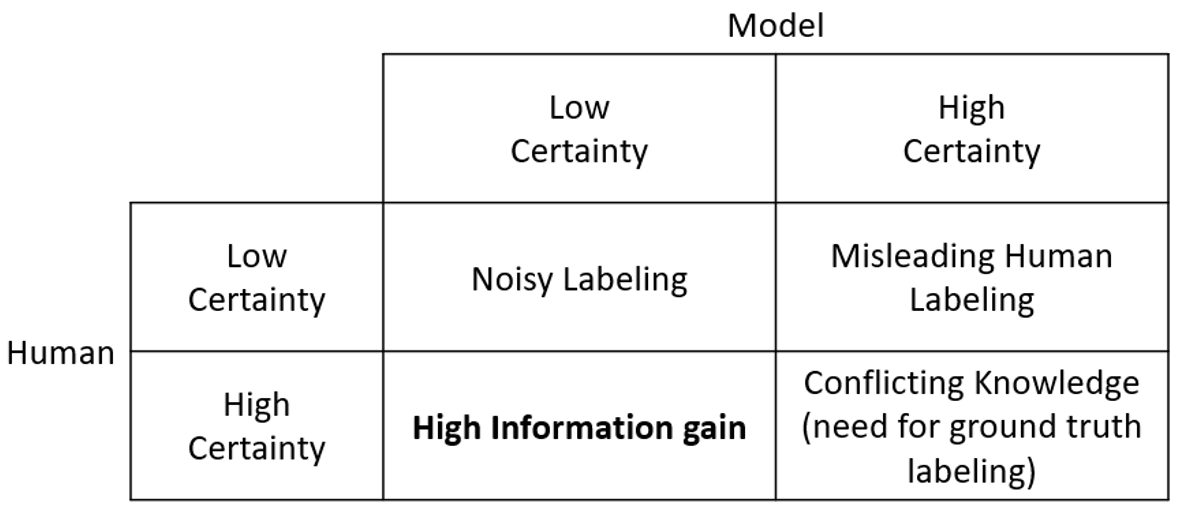}
    \caption{Combinations of human and model certainty and uncertainty (highlighting the “sweet” spot where humans are certain and the model is  uncertain)}
\end{table}

The influence of human uncertainty (from analysts) on model performance can be seen in Figure 8. Each point in the figures represents the average raw confidence level (Y-axis) and the AUPRC increase/decrease in the resulting model in the following round (X-axis). An upward trend (r=0.83; p=0.01) can be seen with the models trained by the individual participant 1 (model Ind\_1) colored in orange; however, other models (Ind\_2, Ind\_3, Swp\_1, Swp\_2, and Grp) combined showed a non-significant or borderline relation between confidence levels and model improvement (r=0.24; p=0.10). Based on these initial results we develop two hypotheses for future research: 1) there is a positive relationship between analyst confidence and model performance when analysts have a sufficient level of labeling expertise; 2) the relationship between rating confidence and resulting model quality is greater for individuals than for groups of raters.

\begin{figure}[h!]
    \centering
    \includegraphics[scale=0.35]{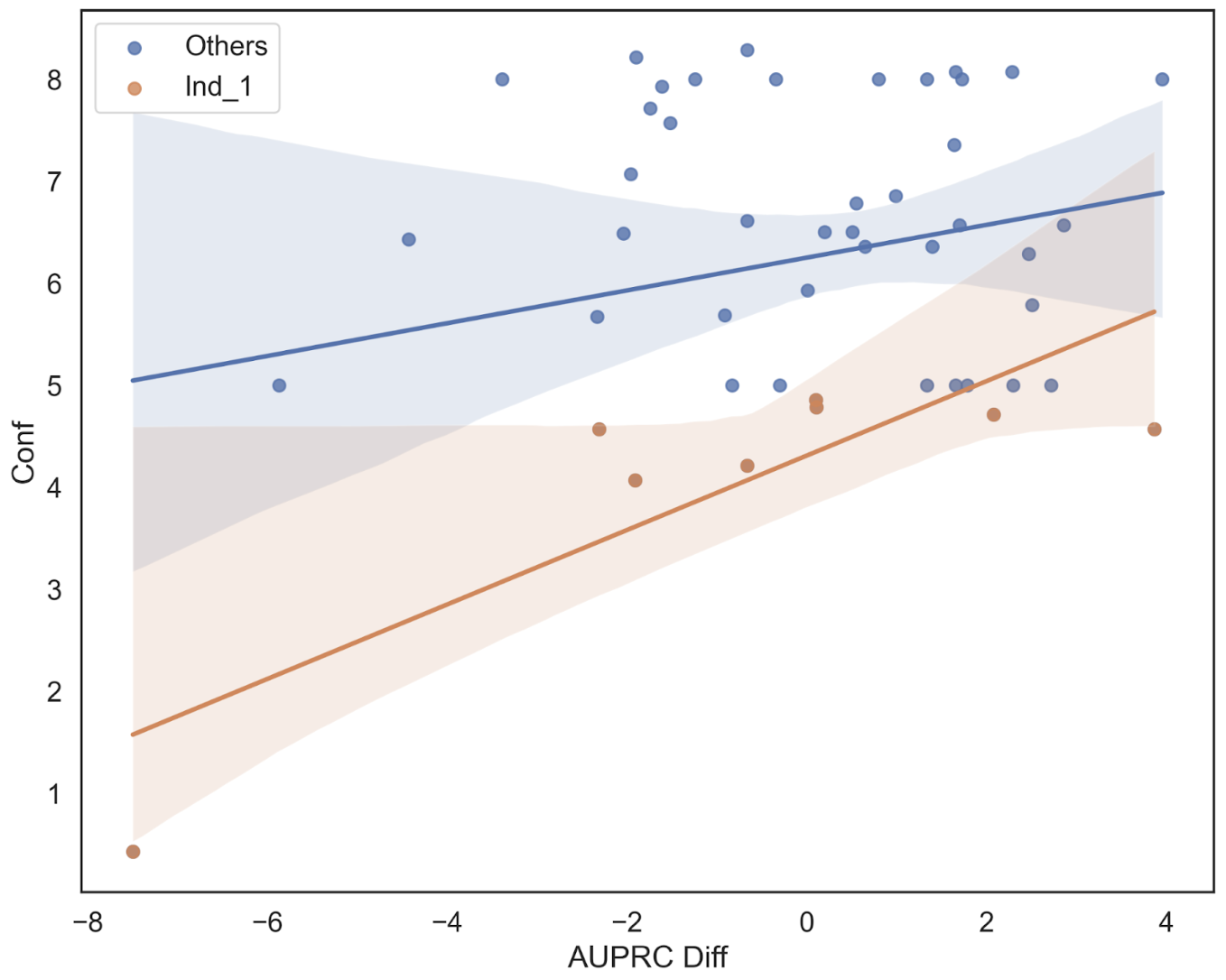}
    \caption{Trends of average raw confidence levels versus model AUPRC change each day of Ind\_1 versus others (Ind\_2, Ind\_3, Swp\_1, Swp\_2, and Grp)}
    \label{}
\end{figure}

\begin{table}[h!]
    \centering
    \includegraphics[scale=0.18]{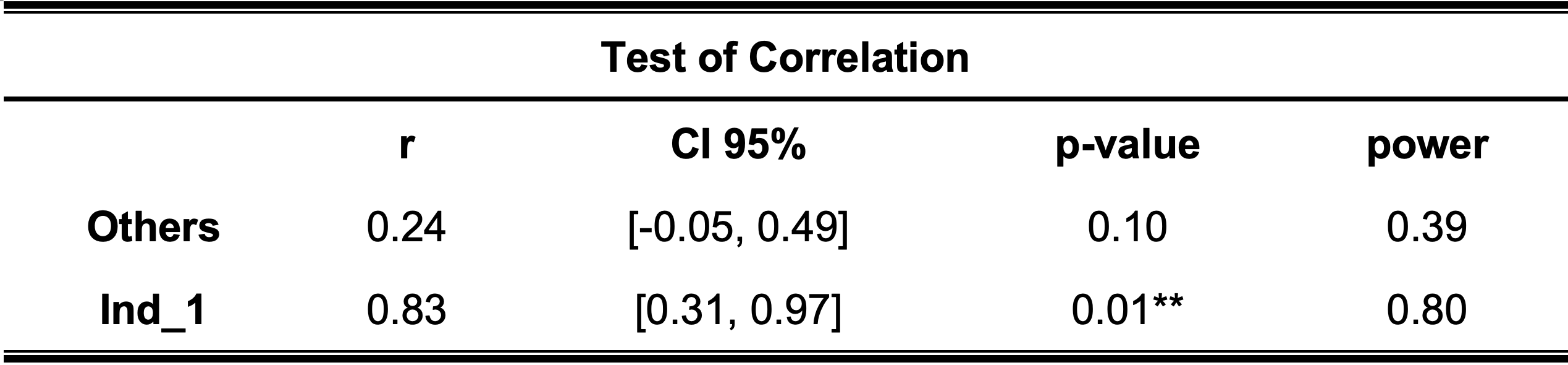}
    \caption{Statistics showing trends in Figure 8 (whether the slope is different from zero)}
\end{table}

In addition, there was a tendency (seen in Figure 9) for high-risk queries (HRQs) to have higher average confidence levels than uncertain queries (UQs). Additionally, it can be seen in Figure 9 that the variations of confidence ratings were much higher in general for the uncertain queries (UQs). This demonstrates that more uncertain cases led to more variation in the confidence of the analysts in their ratings. 

\begin{figure}[h!]
    \centering
    \includegraphics[scale=0.78]{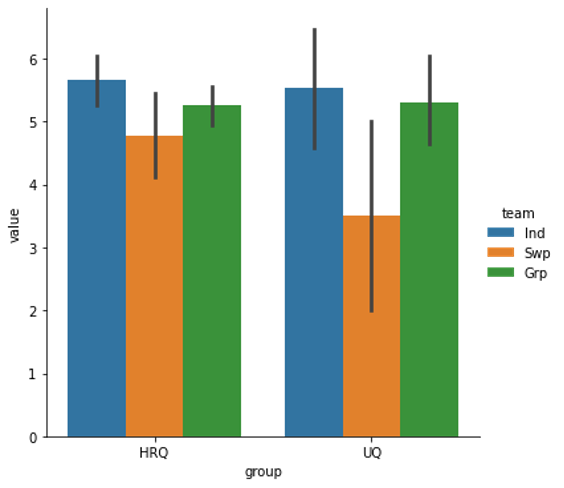}
    \caption{Confidence level differences of HRQs and UQs in terms of their means and variations across participants in the group training team when the label is 1 (True)}
    \label{}
\end{figure}

\subsection{Questionnaire Responses}
At the end of the case study we used a questionnaire to collect feedback regarding analyst experience with the AL models, including both quantitative and qualitative results. Descriptive results for each question are shown in Table 6, followed by a summary and interpretation of the qualitative results. We will then use these qualitative results, as well as the quantitative results presented earlier, to assess the research questions posed earlier in this paper. 

\begin{table}[h!]
    \centering
    \includegraphics[scale=0.48]{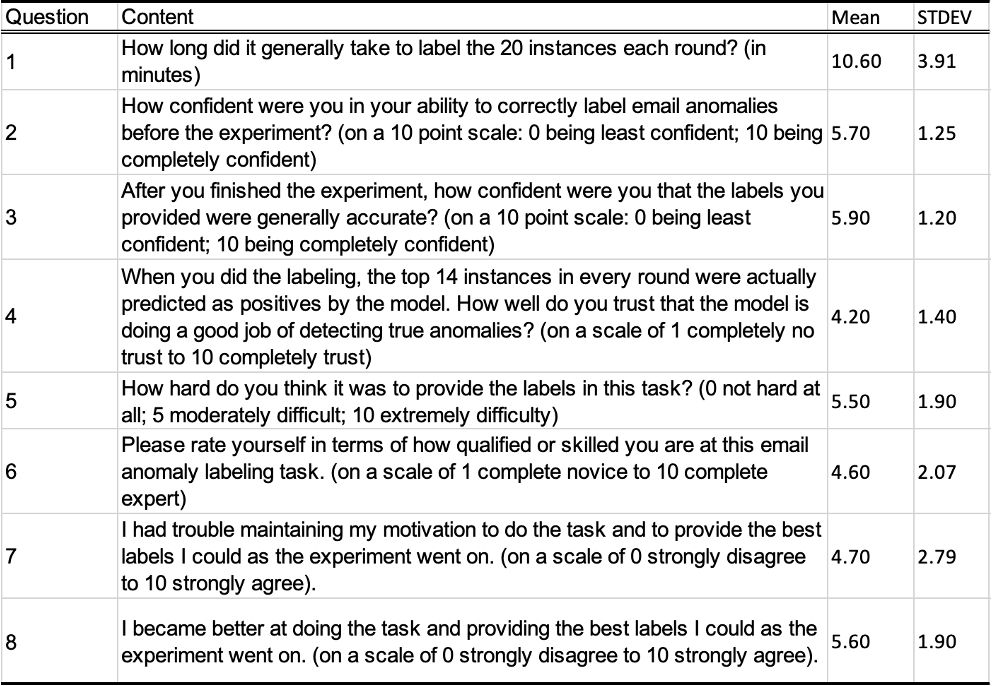}
    \caption{Descriptive statistics (mean and standard deviation) for each of the Questions asked}
    \vspace{-15pt}
\end{table}

Analysts participating in the case study reported that: 
\begin{itemize}
    \item On average it took them around 10 minutes per round to label 20 instances (the process improved from the previous standard which can cost an expert 2 days every week)
    \item They were generally confident about the capability of providing quality labels (9 out of 10 participants have reported confidence levels higher than average, where 4 of them were higher than 2 standard deviations above the mean)
    \item They generally did not find it difficult to maintain motivation (4 out of 10 participants reported difficulty to maintain motivation, where only 1 of them was higher than 2 standard deviations above the mean)
    \item They became better in terms of providing proper labels throughout the experiment (8 out of 10 participants reported they became better, where 4 of them were higher than 2 standard deviations above the mean)
\end{itemize}
 
However, they were concerned about:
\begin{itemize}
    \item The difficulty of labeling (due to the limited amount of information provided for each case)
    \item Whether they were able to trust model outputs 
\end{itemize}

They suggested that the data should include:
\begin{itemize}
    \item Visualization of activity history/comparison within senders across similar roles
    \item More information concerning sensitive/private content (such as attachment details)
\end{itemize}

\subsection{Summary of Findings}
After presenting the model results and questionnaire responses we revisit the research questions proposed in section 1, and interpret the effectiveness of AL in a realistic scenario where ground truth labels were not available. 

\textbf{RQ1}: Are experts well calibrated in terms of assigning self-confidence ratings to their labeling decisions? (i.e., do model predictions improve when experts are more confident in their labeling decisions?)

The uncertainty that analysts have about labeling appeared to strongly affect the ability of AL to improve prediction in the case of one of the participant labelers, who performed best (Figure 8 and Table 5). Thus this research question may be answered in the affirmative in the case of high-performing expert labelers. Ideally, if we can trust human confidence judgments with respect to labeling, then confidence judgments provide a bound on how much improvement can be expected in active learning, since models will only benefit from training with human-supplied labels if the human labels are in fact reliable. Since it is difficult to establish the reliability of human labels when ground truth labels are not available, confidence might be a useful proxy, in cases where participants have sufficient labeling expertise, for the certainty of labeling. 

\textbf{RQ2}: How does the assignment of individuals vs. groups to active learning processes affect subsequent model prediction accuracy?

Labeling by a group of people was found to be less effective than labeling based on individuals. This result contradicts typical findings  in the crowdsourcing (common AL) literature, where participants possess complete knowledge. In these common AL tasks that do not require expertise, redundancy created by multiple annotators can be helpful as a way to improve label quality and reliability. However, in this case study the participants were only provided with partial information about each email. The different confidence/uncertainty levels between participants in the group team (and also the swap team) may have undermined the overall model improvements. , Based on the results we obtained, we suggest that in a corporate environment where expertise is required and privacy is a major concern, well-selected and trained individuals can perform better than groups of  people with varying amounts of experience/knowledge. 

\textbf{RQ3}: How well do experts agree with each other and how reliable are labels across different analysts (i.e., where performance is defined in terms of the prediction metrics of subsequent models)?

Agreement between the participants, as assessed by Krippendorff’s Alpha was relatively low. This does not seem surprising given the difficulty of making labeling decisions in a critical security task, where only partial information is given. Given the strong individual differences that were observed, it seems clear that training sessions are needed to reach a consensus on how to label “correctly”, and to teach participants how to adopt best labeling practices based on the consensus approach identified. Personnel selection may also be needed, so that qualified individuals are recruited for labeling tasks.

\section{DISCUSSION: IMPLICATION AND LIMITATION, AND FUTURE WORK}
The results and findings obtained here provide insights into the challenges of applying expertise in AL and also suggest possible strategies and remedies for dealing with situations where ground truth labels are not available.  Some findings are consistent with previous studies that consider AL tasks in the context of faulty or incomplete human knowledge. Previous researchers have recommended collecting confidence measurements while asking for labels. Combining confidence with dichotomous labels can create multiclass labels, potentially providing more informative and balanced data, and improving model performance.

Some of the present findings differ from earlier studies, (many of which used non-experts in the labeling task). The properties of the cybersecurity domain create a need for novel research strategies. While the use of group training to create redundancy that can be leveraged to filter out incorrect labels (in advance to model teaching) may be feasible in domains where potential experts are plentiful, it is not a feasible strategy in most cybersecurity contexts. Since experts in cybersecurity tend to be scarce and costly resources, the use of large groups for data collection is not feasible. Thus, it may not be feasible to resolve disagreements that may occur between experts by looking for subsets of reliable participants (who agree with each other) within a larger group. 

The present research is relatively unique in looking at Cybersecurity-related AL within a large organization. We hope that the present research will stimulate further “realistic” research in this important area. One downside of conducting research in an industry setting is (as was true for this case study) that it is conducted in an environment with multiple constraints/limitations. For instance, due to privacy concerns and limited security clearance, some captured variables were not accessible in the present case, making the labeling task more uncertain. The limited information available in turn limited the ability of domain experts to use their knowledge fully in providing labels. Another limitation alluded to earlier was the number of participating human participants and their availability for multiple rounds of labeling. In the present study, it took several weeks to get 8 rounds of labeling data even when only 20 labels needed to be assigned in each round. We doubt that this kind of difficulty in collecting data in this kind of setting is atypical. However, the end results of these constraints on the number of iterations (rounds), the size of the query set per round, and the number of expert participants, make it difficult to achieve statistically significant results when addressing some research questions.

The study reported in this paper used 8 rounds of labeling with 20 cases, with 10 participants in each round. While this is a relatively large sample in the context of an industry setting, it is limited in terms of its generalizability. However, it might be argued that AL will likely work differently in different settings, and thus it is more beneficial to look at the impact of confidence ratings, and different groupings of participants with small samples, but in a range of settings, rather than trying to achieve large samples but in a small number of settings. The present results are also useful in showing how AL performs in situations where available data features are constrained due to privacy concerns. 

In spite of the limitations noted above, the findings obtained provide novel insights into the use of AL in a realistic setting and may provide some initial guidance on how to implement AL for cybersecurity applications within organizations. We expect that our study design and the findings obtained should provide a good foundation for future researchers in this area. The problem of interactive ML, or how to get humans to interact with AI is an exceptionally important one that has only started to get due consideration in the past few years. One fertile area for future research involves the relationship between human and model uncertainty. Research is needed to: A) test whether there are better ways to elicit expert knowledge, so as to improve label reliability; B) test whether there are better ways to query experts, such as asking only instances that they are more likely able, or confident, to answer.  

In this study, we did not have access to the amount of time that the analysts used in judging each case during the AL process. However, as noted earlier, the efficiency of AL should ultimately be judged in terms of the tradeoff between model improvement and analyst effort expended. In this paper, we have framed the problem and demonstrated that experts can assign confidence labels that are predictive of subsequent ML prediction accuracy. Since it appears that confidence labeling is useful in guiding AL, future research should examine the relative efficiency of AL methods that are guided by the degree of confidence that human analysts have. 

\section{CONCLUSION}
The characteristics of the cybersecurity domain make it difficult to adopt Machine Learning (ML) technologies without considering domain expert human factors. Domain experts make critical decisions concerning the total security of the organization as well as the related users and customers. Their time is highly valuable and the labels they give are thereby costly to obtain. In this case study we applied Active Learning (AL) as an intervention to a manually conducted email anomaly detection task that was considered difficult for automatic ML. This intervention was intended to improve labeling, training, and detecting efficiencies. We also proposed three research questions to study the human factors of domain analysts and their interactions working with ML models. 

We used a batch-based scenario with a combination of High-Risk Querying (HRQ), Uncertainty Querying (UQ), and Random Querying (RQ) strategies so as to make the labeling task related to the expert daily workload, as well as covering the density and diversity requirements for better AL queries. In  order to study analyst behaviors with AL in different teams, we separated participants into three teams, each with different training schemes: individually training a model;  swapping individually trained models in the middle of the case study; training one model in a group. 

The results provided insights into expert behavior when working with a ML model. Recommendations generated from the findings, and the answers to our research questions, can be summarized as follows: 

\begin{itemize}
    \item Participant confidence levels are crucial because they can train better performing multiclass models
    \item Grouping analysts to create redundancy, and swapping analysts to reduce workload, may not help model performance
    \item Training/selecting guidelines may be needed to find the most appropriate people for carrying out labeling tasks
    \item Better model improvements may be associated with higher average confidence levels, and labelers should be selected based on the labeling expertise so that models are trained better and perform better
    \item In designing labeling procedures, organizations should consider the possible trade-off between querying model uncertain instances to obtain more information gain, and querying in- stances that human analysts are confident in to obtain better quality labels
\end{itemize}

Based on this research, we recommend that AL algorithms should be able to either a) predict expert confidence level and only request labels with instances that experts can confidently make decisions; or b) accept multilabel or multiclass labels so that instances that humans are uncertain about labeling can be filtered out without hurting model performance. 

In conclusion, there is an urgent need to improve the effectiveness of defenses against cybersecurity threats. While great advances have been made in areas such as perimeter defense and the use of ML to detect anomalies, the role of human expertise in aiding ML performance has often been overlooked. While techniques such as AL provide a way to incorporate human expertise into the critical label- ing/training task in ML, the present results also demonstrate the complexities that arise in incorporating human expertise. Three major issues highlighted in the research reported here are: 1) How to incorporate the uncertainty that experts have in the labels they are assigning; 2) How to deal with individual differences in labeling expertise; 3) How best to use multiple experts (in terms of grouping and refining their judgments) during AL.

\vspace{12pt}

\end{document}